# Causal relationships between eruptive prominences and coronal mass ejections

B. Filippov[1] and S. Koutchmy[2]

[1]{Institute of Terrestrial Magnetism, Ionosphere and Radio Wave Propagation, Russian Academy of Sciences (IZMIRAN), Troitsk Moscow Region, Russia }

[2]{Institut d'Astrophysique de Paris, CNRS and Univ. P.& M. Curie, Paris, France}

Correspondence to: B. Filippov (bfilip@izmiran.ru)

**Abstract**

A close association between eruptive prominences and CMEs, both slow and fast CMEs, was reported in many studies. Sometimes it was possible to follow the material motion starting from the prominence (filament) activation to the CME in the high corona. Remnants of the prominence were found in the bright core of the CME. However, detailed comparisons of the two phenomena reveal problems in explaining CMEs as a continuation of filament eruptions in the upper corona. For example, the heliolatitudes of the disappeared filaments and subsequent coronal ejections sometimes differ by tens of degrees. In order to clear up the problems appearing when considering this association EP- CME, we tentatively analyse the more general question of the dynamics of the generic magnetic flux rope. Prominences and filaments are *the best* tracers of the flux ropes in the corona long before the beginning of the eruption. A twisted flux rope is held by the tension of field lines of photospheric sources until parameters of the system reach critical values and a catastrophe happens. We suggest that the associated flux rope height above the photosphere is one of these parameters and that it is revealed by the measured height of the filament. 80 filaments were analysed and we found that eruptive prominences were near the so-called limit of stability a few days before their eruptions. We suggest that a comparison of actual heights of prominences with the calculated critical heights from magnetograms could be systematically used to predict filament eruptions and the corresponding CMEs.

# 1  Introduction

Hα observations of the activation of a filament (prominence) frequently end up suddenly due to the shift of this line from the filter pass-band due to i) the Doppler effect, ii) the decrease of the emission due to the decrease in the density of expanding material, and iii) the ionization of hydrogen due to the heating of the prominence plasma. The subsequent evolution can sometimes be observed in ultraviolet or radio, but the fate of a prominence usually becomes visible only when it reaches the lower boundary of the field of view of a coronagraph and becomes visible in white light as a coronal mass ejection (CME). A typical CME consists of three parts: a bright core that is the remnant of an eruptive prominence (EP); a large, dark, lower-density surrounding cavity; and an outer, rather diffuse envelope having the projected shape of a closed loop with its legs fixed on the Sun (Crifo et al, 1983; Sime et al., 1984).

There is a very close association between eruptive prominences and CMEs. There are no doubts, that in ejections with bright cores, the material of eruptive prominences is contained in the cores (House et al, 1981). Figure 1 shows an example of event when an eruptive prominence was recorded in He II 304 Å line with the ultraviolet telescope EIT on SOHO, and then the remnants of the twisted prominence were well recognised in the core of the CME up to a distance of several solar radii from the surface. Another example (Fig. 2) indicates a good correspondence in time, spatial position and direction of motion of an EP filament and a CME. Even in the case of a limb CME observed at the time of a solar total eclipse, when the inner and the outer white-light corona are both well imaged, this association is obvious (see Koutchmy et al., 2004) including the occurrence of a high loop/cavity association. At the same time, there are ejections without bright core, or events, for which it is impossible to find the presumed prominence eruption or filament disappearance. The opposite, when a CME is absent after a filament eruption, was reported. However this happens in the case of "confined" eruptions, when the filament does not fly away too far but stops at some height in a new equilibrium position (Vrsnak et al., 1990; Filippov and Koutchmy, 2002; Torok and Kliem, 2005; Alexander et al., 2006). Thus, the relationship between these two phenomena is not always obvious, and it is the subject of analysis in many works.

The purpose of this paper is to discuss these and some other problems of EP-CME association and try to resolve them in the frame of a magnetic flux rope model.

## 2  EPs and CMEs temporal and spatial distributions

Among all manifestations of activity in the lower layers of the solar atmosphere, statistical studies showed the greatest correlation of CMEs with eruptive prominences (Webb et al., 1976; Munro et al, 1979; Webb and Hundhausen, 1987). An important factor is the height at which the eruptive prominence is observed: the greater is the height it reaches, the more probable a CME will follow. So, all prominences studied by Munro et al. (1979) that reached height of $1.2R_\odot$ and 60% of prominences that reached the height $1.1R_\odot$ were associated with CMEs. This raises the question of whether or not the prominences, that at first show a sufficiently rapid ascending motion holding its initial shape and then slow down and stop demonstrating a finite motion, could be called eruptive. Gilbert et al. (2000) proposed to call such prominences active prominences. On the basis of observations in 1996-1998, they found that CMEs were connected to 46% of active prominences and to 94% of "real" eruptive prominences, capable to overcome the solar gravity. As a rule, ejections associated with eruptive prominences had bright cores, while CMEs associated with the so-called active prominences usually did not.

The choice of data used in the correlation analysis can probably influence the result. For example, a low correlation of 10-30% between CMEs and prominence eruptions together with sudden filament disappearances was obtained by Yang and Wang (2002); it is possibly a consequence of including thermal disappearances of filaments not connected with dynamic events into the data set (Mouradian et al., 1995). In the same period of time, 82% of eruptive prominences observed with the Nobeyama radioheliograph were associated with CMEs (Gopalswamy et al, 2003). However, only 18% of the prominences that move predominantly parallel to the surface of the Sun ("transverse" events), and can apparently be assigned to active prominences, were possibly associated to CMEs.

A careful comparison of the positions of filaments that suddenly disappeared and the apparently associated CMEs reveals some problems in the interpretation of CMEs as continuations of the filament eruptions in the upper corona. The helio-latitudes of the disrupted filaments and of the subsequent coronal ejections sometimes differ by tens of degrees. For example, Cliver and Webb (1998) found that, in the period of maximum activity, the number of CMEs with heliolatitudes > 60° was four times the number of disrupted filaments at latitudes > 45°.

Figure 3 shows the distribution of eruptive prominences and CMEs about the magnetic equator for the epoch of solar minimum. Data on disappearing filaments and eruptive prominences were taken from the catalog of solar events presented by the NOAA at *http://www.sec.noaa.gov/ftpmenu/indices/events.html*. CME positions were taken from SOHO/LASCO observations. The neutral line from maps of the magnetic field of the source surface calculated and published monthly in the Solar Geophysical Data was used as the solar magnetic equator. Prominence eruptions take place fairly far from the plane of the magnetic equator, while CMEs tend to concentrate toward this plane. Therefore, if indeed a CME represents a further development of an eruptive prominence, the latter would have to travel a long distance along the solar limb rather than rise radially (Westin and Liszka, 1970; Filippov et al., 2001). An example of such non-radial eruptive prominence motion is presented in Fig. 4. Note that the CME moves radially and along the equatorial plane. An other interesting case of a slow W-limb high latitude CME was analyzed by Boulade et al. (1997), showing a connection with equatorial structures.

A further objection against the concept of a CME as a further development of the eruption of a prominence in the upper corona is based on the kinematics of these phenomena. The velocity of an eruptive prominence (bright core of a CME) is usually less than the velocity of the outer loop (envelope of the CME) by a factor as large as 1.5 (Maricic et al., 2004; Foley et al., 2003). For this reason, it is believed that the prominence could not "push" the CME and be its driver.

Back-extrapolation of the time dependence of the CME height allows the determination of the onset time of the CME. Although some uncertainty remains because of the absence of information about the real CME acceleration during a period of time when it is shielded by the occulting disk of the coronagraph, which makes unknown the starting height, nevertheless it is possible to consider that the onset time of a CME coincides rather well (within half an hour) with the beginning of the prominence eruption (Gopalswamy et al., 2003).

The question: which of the two phenomena, filament eruption or CME, starts earlier? is important for understanding the causal relationship between them. That a rising prominence disturbs the surrounding corona and like a piston pushes out coronal plasma forming a CME is a possible point of view; there is the opposite approach when the prior removal of the coronal structures above a prominence is assumed to allow it to rise unimpeded.

Observations cannot yet give an unambiguous answer to this question, which results from the absence of any measurements of the surrounding magnetic field. More and more arguments are gathered in favour of a concept leading to both prominence eruptions and CMEs being different observational manifestations of the same common process of loss of equilibrium of a large-scale coronal current system (magnetic flux rope). Depending on particular initial conditions in the system (density and mass of the prominence, magnetic field strength, etc.) the pattern of event can be different. In addition, some eruption of very rarefied prominences can remain unnoticed and cannot create a sufficiently bright core of a CME.

Strictly speaking, all three well recognized eruptive phenomena in the solar atmosphere - flares, eruptive prominences and coronal mass ejections - are closely related. The degree of correlation between them depends on special features of a particular phenomenon (Lin, 2004). The correlation between CMEs and flares depends on the quantity of energy stored in the magnetic configuration prior to eruption. The more free energy is stored, the better is the correlation. On the other hand, the correlation between CMEs and eruptive prominences depends on the amount of matter contained in the initial magnetic structure.

## 3  The flux rope model and prediction of eruptive phenomena

Most of abovementioned problems can be resolved in the context of a magnetic flux rope model. Twisted magnetic structures are embedded into the corona in equilibrium under the action of different forces like magnetic pressure created by electric current loops and the diamagnetic properties of the photosphere, the magnetic tension of field lines anchored in the photosphere, and so on (e.g., Mouschovias and Poland, 1978; Chen, 1989; Vrsnak, 1990). Most of the flux rope volume is filled with depleted coronal material and forms a coronal cavity (Fig. 5). Recent multi-wavelength observations were reported, for example, by Koutchmy et al. (2007) and models of density depletion within coronal flux ropes were developed in works of Gibson and Low (1998), Krall and Chen (2005). Helical magnetic flux tubes of the rope serve as magnetic traps in the gravity field. Dense plasma can be collected near the bottom parts of the helical field lines forming a prominence (Demoulin and Priest, 1989). We observe plasma thanks to emitted or absorbed radiations. Filaments are the most accessible indicators of the flux ropes presence in the corona but we recall that no direct magnetic field measurements are possible in the corona. Loss of the flux rope equilibrium appears first as a filament eruption and then as a coronal mass ejection. Rising into the upper corona, the cross-section of the flux rope significantly increases. The top part of the flux rope then moves faster than the central and bottom parts. That is why the CME's front has usually a higher velocity than the eruptive prominence. A selfsimilar dynamical evolution of a partly anchored, twisted magnetic flux rope was presented by Low and Hundhausen (1987), Gibson and Low (1998).

Neglecting inertial forces, the flux rope will move along a neutral surface, the surface passing through the apices of the magnetic arches. Figure 6 presents an axially-symmetric flux rope model with a global magnetic field typical for minimum activity (Filippov et al., 2001; 2002) with the quasi-static trajectory shown by a green line. The rope initially moves in a non-radial direction and then radially, as it is presented in Fig. 5. Such behavior could explain the difference in angular distribution of EPs and CMEs shown in Fig. 3.

The non-linear properties of the flux rope model leads to a catastrophic loss of equilibrium when the electric current strength reaches a critical value (Van Tend and Kuperus, 1978; Molodensky and Filippov, 1987; Vrsnak, 1990; Forbes and Isenberg, 1991; Lin et al, 1998; Filippov and Den, 2001). Unfortunately, not only electric currents but even magnetic fields cannot be measured in the corona. At the same time, the model of the flux rope equilibrium shows that the greater is the electric current, the higher is the equilibrium point. Therefore, the height of a filament $h_p$ could be in some sense a measure of the electric current strength and there is a critical height $h_c$ for the stable filament equilibrium. The critical height characterises the scale of the magnetic field and can be found from the equation

$$h_c = \frac{B}{dB/dh\big|_{h_c}} ,$$

using photospheric magnetic field measurements and a current-free approximation (Filippov and Den, 2000; 2001). It corresponds to the critical strength of the flux rope electric current but, in the framework of a simple model, depends only on the background magnetic field distribution. Above the critical height the photospheric magnetic field cannot hold any electric current in stable equilibrium.

Comparing the observed height of filaments with the parameter $h_c$, it is possible to work on predictions of filament eruption occurrence. The height of a filament above the photosphere is not always easy to measure when it is projected against the solar disk. When simultaneous chromospheric images are taken at different angles, as the STEREO mission does for the corona, is an ideal case to determine the filament height from the parallax angles. Sometimes, when the internal motion is negligible within the filament and the filament shape is invariable, the filament height can be estimated from the difference between the shift of chromospheric details and the shift of the top of the filament produced by solar rotation (Vrsnak et al., 1999). One can also use information on the inclination of the symmetry plane of a filament with respect to the vertical in order to estimate the filament's height from its observed width (Zagnetko et al., 2005).

Figure 7 shows the distribution of about 80 filaments observed according to their actual height and the calculated for the surrounding magnetic field critical height. The height $h_p$ of some filaments was measured on the east limb as the height of prominences above the limb using Hα filtergrams of the Big Bear Solar Observatory and Ca II $K_3$ spectroheliograms of the Meudon Observatory. These data were taken from our early work (Filippov and Den, 2001}. However, the height of the most part of filaments was measured on the disk using the above-mentioned methods and Hα filtergrams obtained by the Global High Resolution Hα Network. Potential magnetic fields were calculated on the basis of SOHO/MDI magnetograms. The bisector corresponds to the limit of stability. Quiescent filaments more or less evenly fill the angle between the bisector and the horizontal axis, while the eruptive filaments tend to cluster about the bisector. This shows that eruptive prominences were near the limit of stability a few days before eruptions. Plots in Fig. 8 show some examples of the temporal behavior of filament heights and critical heights for several filaments observed in the latter part of 2005. Right ends of the curves correspond to the last moments just before a filament erupts or disappears from the disk. As a rule it is seen that the filament erupts soon (a day or two) after it reaches the critical height. However, there are exceptions, possibly due to errors in measurements or to limitation of the model.

Chen and Krall (2003), Chen et al. (2006), Krall and Sterling (2007) introduced different critical heights $Z_0$ and $Z_m$ related to the footpoint separation distance $S_f$ for an eruptive flux rope. These parameters limit the height $Z_{max}$ where the acceleration of the apex of the flux rope is maximum. The critical height $h_c$ is the scale height of the radial variation in the local potential field, so it is a property of the external magnetic field, while $Z_0$ is the radius of the semi-circular flux rope and, therefore, it is defined by the geometry of the flux rope. The height $h_c$ is the start point of flux rope acceleration, whereas $Z_{max}$ is the height where the acceleration reaches its peak value. The parameters $Z_0$ and $Z_m$ concern the dynamical behavior of the eruptive flux rope but from what we understand, they do not clearly define the equilibrium conditions and the stability threshold. We believe it is interesting to consider these questions in future works.

## 4  Conclusions

In a number of cases a detailed correlation of prominence eruptions and coronal mass ejections permits to trace the continuous transformation of one phenomenon into another, which makes it possible to speak about a common eruptive phenomenon. A CME is certainly formed in a much more rarefied medium and occupies an enormous volume. Significant parts of the corona and the large-scale magnetic field are involved in the CME formation; however, the prominence material is often well distinguished in the general structure of the CME and its bright core. Thus, the problem of the sudden ejection of plasma from the corona into the interplanetary space and the problem of the prominence existence and eruption seems to be quite closely connected and their solutions should be searched for in a general context.

The filament height can be a good indicator of its lifetime duration and, accordingly, of the probability to have the filament erupting. The comparison of the real heights of prominences with the calculated critical heights could be a basis for predicting the filament eruptions and the associated CMEs. Solar storm forecast several days in advance could be made in such events on the basis of the filament monitoring. All necessary data could be obtained, in principle, with ground-based instruments, although routine chromospheric space-born observations could significantly increase the precision.

**Acknowledgements**

The authors are grateful to the SOHO20 organisers for the financial support given to attend the meeting. This work was supported in part by the Russian Foundation for Basic Research (grant 06-02-16424).

# References

Alexander, D., Liu, R., and Gilbert, H.: Hard X-ray production in a failed filament eruption, Astrophys. J., 653, 719-724, 2006.

Boulade, S., Delannée, C., Koutchmy, S., Lamy, P., Llebaria, A., Howard, R., Schwenn, R., and Simnett, G.: Analysis of a high latitude slow CME with travelling ejecta, The Corona and Solar Wind Near Minimum Activity, A. Wilson (Ed.), ESA SP, 404, 217-221, 1997.

Chen, J.: Effects of toroidal forces in current loops embedded in a background plasma, Astrophys. J., 338, 453-470, 1989.

Chen, J., and Krall, J.: Acceleration of coronal mass ejections, J. Geophys. Res., 108 (A11), 1410, doi:10.1029/2003JA009849.

Chen, J., Marque, C., Vourlidas, A., Krall, J., and Schick, P. W.: The flux-rope scaling of the acceleration of coronal mass ejections and eruptive prominences, Astrophys. J., 649, 452-463, 2006.

Cliver, E. W. and Webb, D. F.: Disappearances of high-latitude filaments as sources of high-latitude CMEs, New Perspectives on Solar Prominences, Webb, D., Rust, D., and Schmieder, B., (Eds.), ASP Conference Series; ASP: San Francisco, CA, Vol. 150, 479-483, 1998.

Crifo, F., Picat, J. P., and Cailloux, M.: Coronal transients - loop or bubble, Solar. Phys., 83, 143-152, 1983.

Demoulin, P. and Priest, E. R.: A twisted flux model for solar prominences. II - Formation of a dip in a magnetic structure before the formation of a solar prominence, Astron. Astrophys., 214, 360-368. 1989.

House, L. L., Wagner, W. J., Hildner, E., Sawyer, C., and Schmidt, H. U.: Studies of the corona with the Solar Maximum Mission coronagraph/polarimeter, Astrophys. J., 244, L117-L121, 1981.

Filippov, B. P. and Den, O.G.: Prominence height and vertical gradient in magnetic field, Astron. Lett., 26, 322-327, 2000.

Filippov, B. P. and Den, O. G.: A critical height of quiescent prominences before eruption. J. Geophys. Res., 106, 25177-25184, 2001.

Filippov, B. P., Gopalswamy, N., and Lozhechkin, A. V.: Non-radial motion of eruptive filaments, Solar Phys., 203, 119-130, 2001.

Filippov, B. P., Gopalswamy, N., and Lozhechkin, A. V.: Motion of an eruptive prominence in the solar corona, Astron. Reports., 46, P. 417-422, 2002.

Filippov, B. P. and Koutchmy, S.: About the prominence heating mechanisms during its eruptive phase, Solar Phys., 208, 283-295, 2002.

Foley, C. R., Harra, L. K., Matthews, S. A., Culhane, J. L., and Kitai, R.: Evidence for a flux rope driven EUV wave and CME: comparison with the piston shock model, Astron. Astrophys., 399, 749-754, 2003.

Forbes, T. G. and Isenberg, P. A.: A catastrophe mechanism for coronal mass ejections, Astrophys. J., 373, 294-307, 1991.

Gibson, S. E. and Low, B. C.: A time-dependent three-dimensional magnetohydrodynamic model of the coronal mass ejection, Astrophys. J., 493, 460-473, 1998.

Gilbert, H. R., Holzer, T. E., Burkepile, J. T., and Hunhausen, A. J.: Active and eruptive prominences and their relationship to coronal mass ejections, Astrophys. J., 537, 503-515, 2000.

Koutchmy, S., Baudin, F., Bocchialini, K., Daniel, J.-Y., Delaboudinière, J.-P., Golub, L., Lamy, P., and Adjabshirizadeh, A.: The August 11th, 1999 CME, Astron. Astrophys., 420, 709-718, 2004.

Koutchmy, S., Filippov, B., and Lamy, Ph.: Old and new aspects of prominence physics from coronal observations, The Physics of Chromospheric Plasmas, Heinzel, P., Dorotovič, I., and Rutten, R. J., (Eds.), ASP Conference Series, ASP: San Francisco, CA, Vol. 368, 331-336, 2007.

Krall, J. and Chen, J.: Density structure of a preeruption coronal flux rope, Astrophys. J., 628, 1046-1055, 2005.

Krall, J. and Sterling, A. C.: Analysis of erupting solar prominences in terms of an underlying flux-rope configuration, Astrophys. J., 663, 1354-1362, 2007.

Kuperus, M. and Raadu, M. A.: The support of prominences formed in neutral sheets, Astron. Astrophys., 31, 189-193, 1974.

Lin, J.: CME-flare association deduced from catastrophic model of CMEs, Solar Phys., 219, 169-196, 2004.

Lin, J., Forbes, T. G., Isenberg, P. A., and Demoulin, P.: The effect of curvature on flux-rope models of coronal mass ejections, Astrophys. J., 504, 1006-1019, 1998.

Low, B. C. and Hundhausen, A. J.: The velocity field of a coronal mass ejection - The event of September 1, 1980, J. Geophys. Res., 92, 2221-2230, 1987.

Maričić, D., Vršnak, B., Stanger, A. L., and Veronig, A.: Coronal mass ejection of 15 May 2001: I. Evolution of morphological features of the eruption, Solar Phys., 225, 337-353, 2004.

Molodensky, M. M. and Filippov, B. P.: Rapid Motion of Filaments in Solar Active Regions. II, Soviet Astron., 31, 564-568, 1987.

Mouradian, Z., Soru-Escaut, I., and Pojoga, S.: On the two classes of filament-prominence disappearance and their relation to coronal mass ejections, Solar Phys., 158, 269-281, 1995.

Mouschovias, T. Ch. and Poland, A. I.: Expansion and broadening of coronal loop transients - A theoretical explanation, Astrophys. J., 220, 675-682, 1978.

Sime, D. G., MacQueen, R. M., and Hundhausen, A. J.: Density distribution in looplike coronal transients - A comparison of observations and a theoretical model, J. Geophys. Res., 89, 2113-2121, 1984.

Munro, R. H., Cosling, J. T., Hildner, E., Mac Queen, R. M., Poland, A. I., and Ross, C. L.: the association of coronal mass ejection transients with other forms of solar activity, Solar Phys., 61, 201-215, 1979.

Gopalswamy, N., Shimojo, M., Lu, W., Yashiro, S., Shibasaki, K., and Howard, R. A.: Prominence eruptions and coronal mass ejection: a statistical study using microwave observations, Astrophys. J., 586, 562-578, 2003.

Torok, T. and Kliem, B.: Confined and ejective eruptions of kink-unstable flux ropes, Astrophys. J., 630, L97-L100, 2005.

Van Tend, W. and Kuperus, M.: The development of coronal electric current system in active regions and their relation to filaments and flares, Solar Phys., 59, 115-127, 1978.

Vrsnak, B.: Eruptive instability of cylindrical prominences, Solar Phys., 129, 295-312, 1990.

Vrsnak, B., Ruzdjak, V., Brajsa, R., and Zloch, F.: Oscillatory motions in an active prominence, Solar Phys., 127, 119-126, 1990.

Vrsnak, B., Rosa, D., Bozic, H. et al.: Height of tracers and the correction of the measured solar synodic rotation rate: demonstration of the method, Solar Phys., 185, 207-225, 1999.

Webb, D. F., Krieger, A. S., and Rust, D. M.: Coronal X-ray enhancements associated with H-alpha filament disappearances, Solar Phys., 48, 159-186, 1976.

Webb, D. F., Hundhausen, A. J.: Activity associated with the solar origin of coronal mass ejections, Solar Phys., 108, 383-401, 1987.

Westin, H. and Liszka, L.: Motion of ascending prominences, Solar Phys., 11, 409-424, 1970.

Yang, G. and Wang, Y.: Statistical Studies of Filament Disappearances and CMEs, Magnetic Activity and Space Environment; Wang, H. N., Xu, R. L. (Eds.), Proc. COSPAR Colloq.; Pergamon: Boston, Ser. 14, 113, 2002.

Zagnetko, A. M., Filippov, B. P., and Den O. G.: Geometry of solar prominences and magnetic fields in the corona, Astron. Reports, 49, 425-430, 2005.

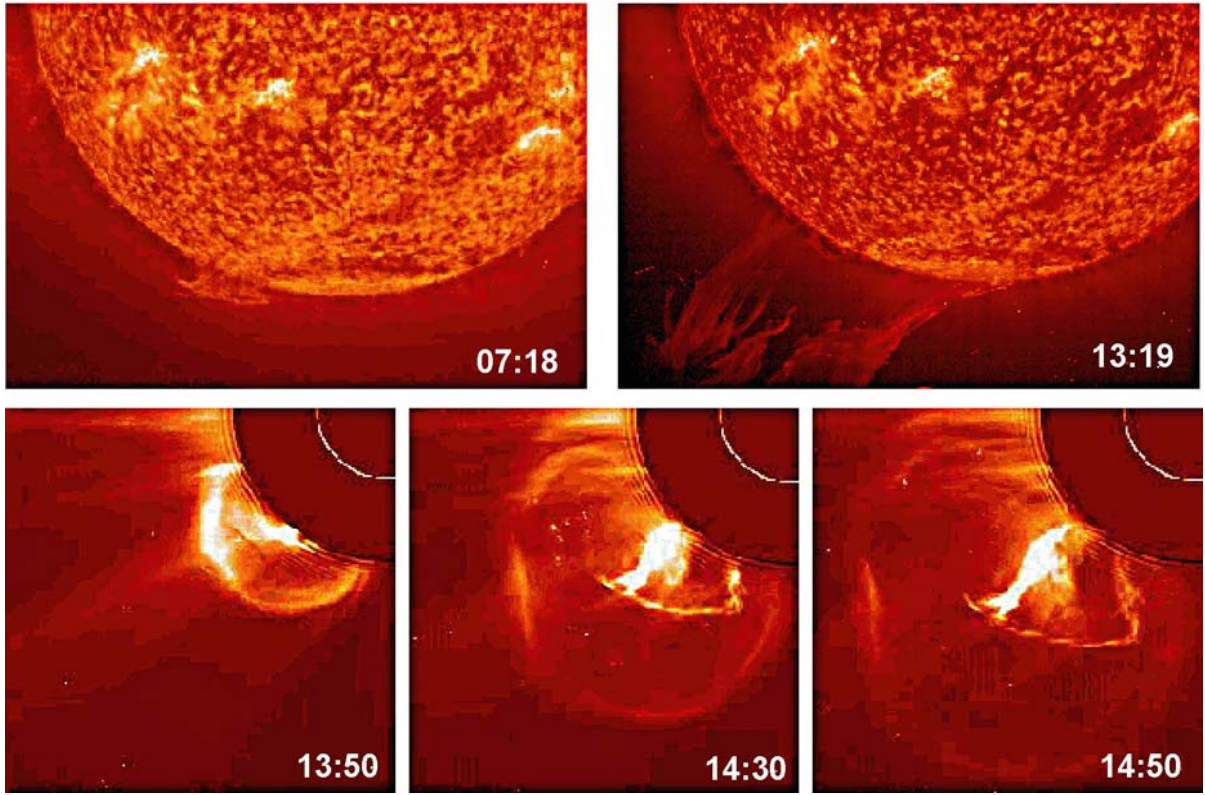

Figure 1. Polar crown filament eruption on 14 June 1999 visible in SOHO/EIT He II 304 Å line (top row) and the following CME observed with LASCO C2 (bottom row). (Courtesy of SOHO/EIT and SOHO/LASCO consortiums. SOHO is a joint ESA-NASA project.)

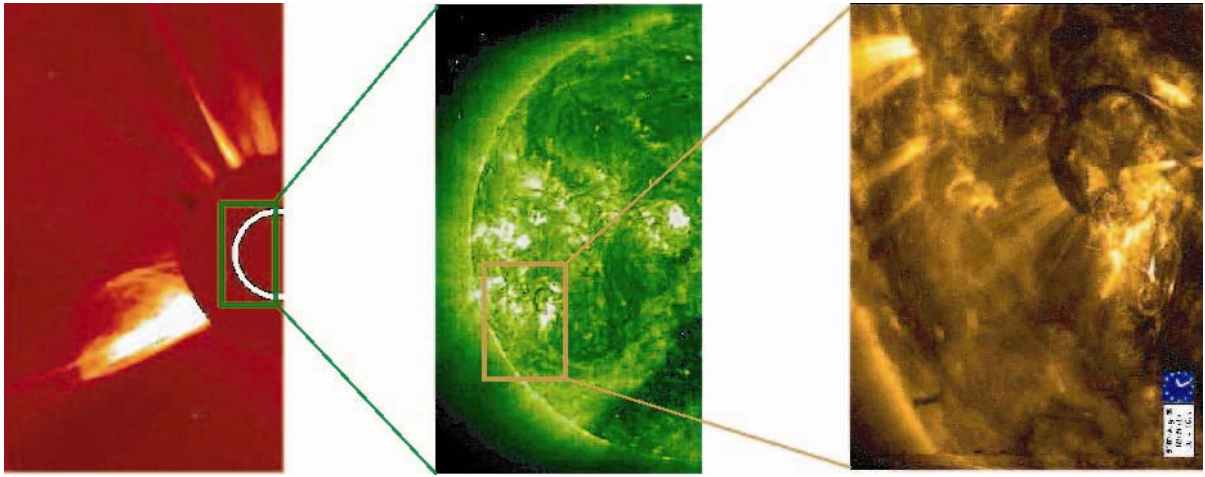

Figure 2. Eruptive process on 25 August 2003 shown in subsequently increasing scale. White light SOHO/LASCO C2 corona at 04:26 UT (left), SOHO/EIT Fe XII 195 Å line image at 02:24 UT (middle), and TRACE Fe IX 171 Å line image at 02:26 UT (right). Eruptive filament just after the start of ascending motion is seen as dark semicircular feature at the upper-right corner of the TRACE filtergram. (Courtesy of SOHO/EIT, SOHO/LASCO, and TRACE consortiums. SOHO is a joint ESA-NASA project. TRACE is a mission of the Stanford-Lockheed Institute for Space Research (a joint program of the Lockheed-Martin Advanced Technology Center's Solar and Astrophysics Laboratory and Stanford's Solar Observatories Group) and part of the NASA Small Explorer program.))

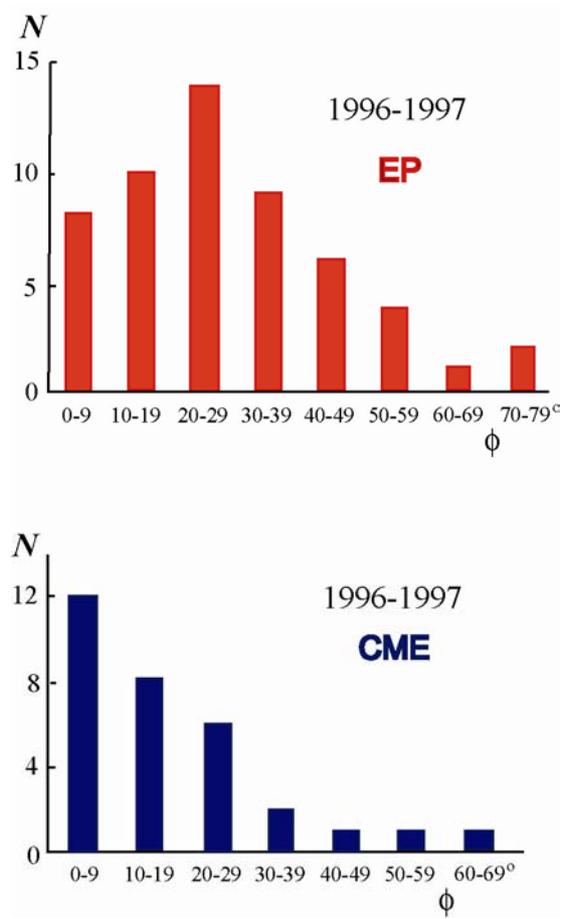

Figure 3. Angular distribution of eruptive prominences and CMEs about the magnetic equator ($\varphi = 0$) for the epoch of solar minimum.

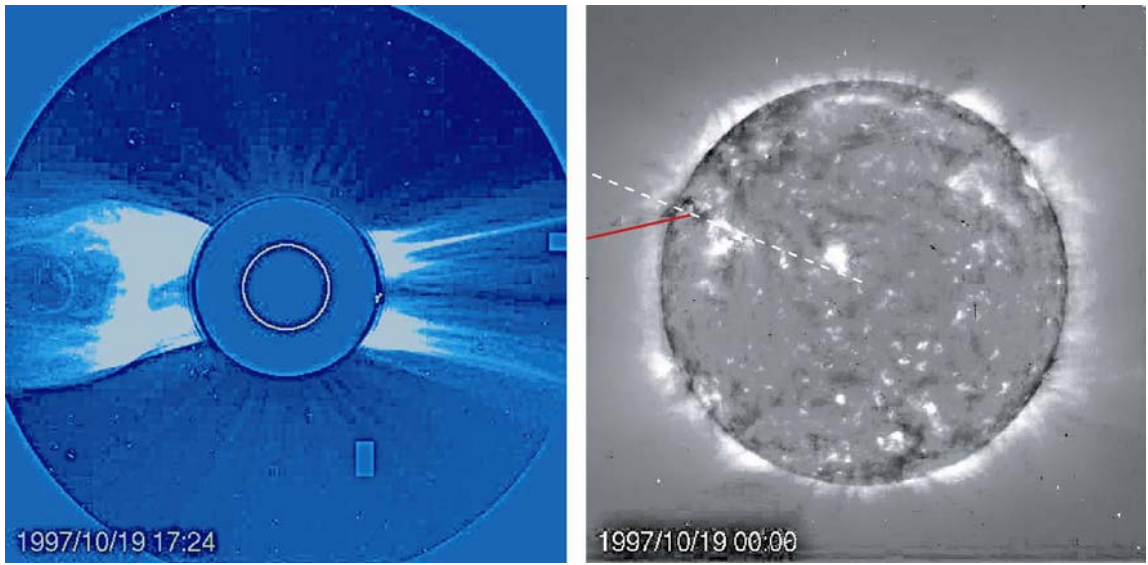

Figure 4. CME on 19 October 1997 in a field of view of SOHO/LASCO C2 (left) and composite differential image obtained from a series of SOHO/EIT Fe XII 195 Å line images (right). A chain of dark and bright nodes above the red line traces the path of the eruptive prominence. The trajectory deviates more than 30° southward from the radial direction, shown with the white dashed line. The shape of the CME is nearly symmetric about the equatorial plane and it moves radially. (Courtesy of SOHO/EIT and SOHO/LASCO consortiums. SOHO is a joint ESA-NASA project.)

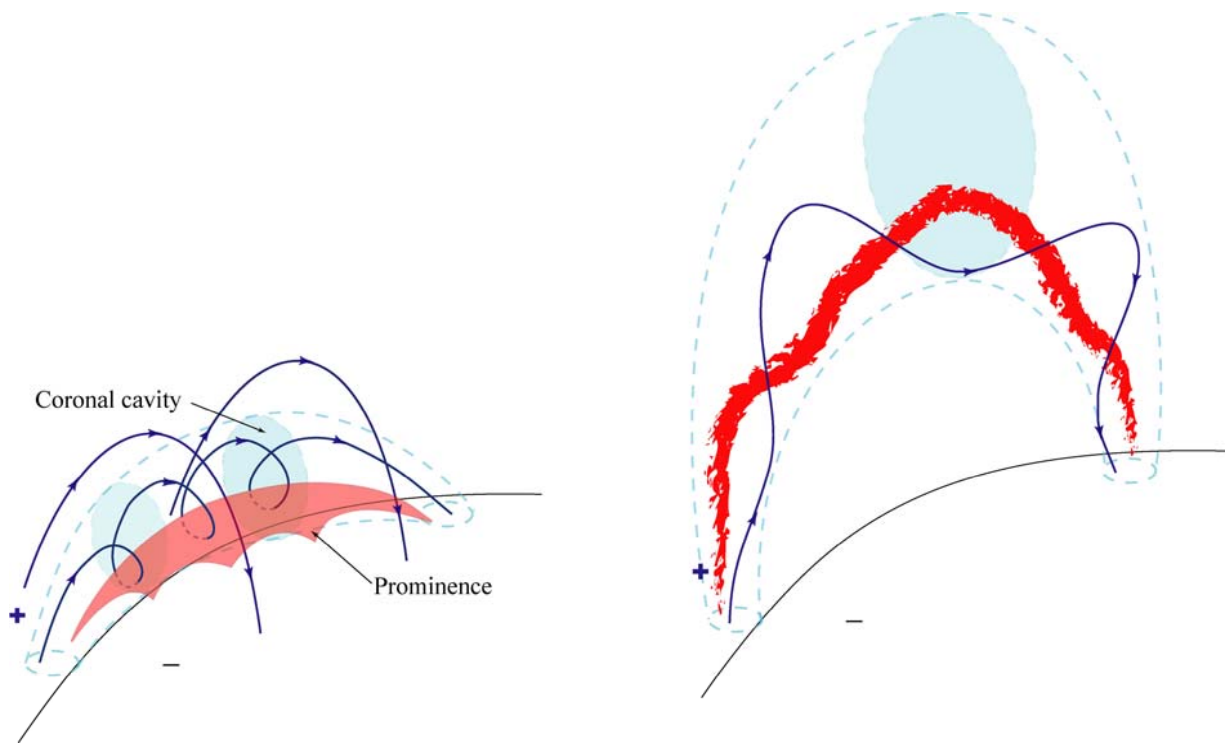

Figure 5. A sketch of the 3D geometry of a magnetic flux rope with a coronal cavity and a prominence inside in equilibrium (left) and during eruption (right).

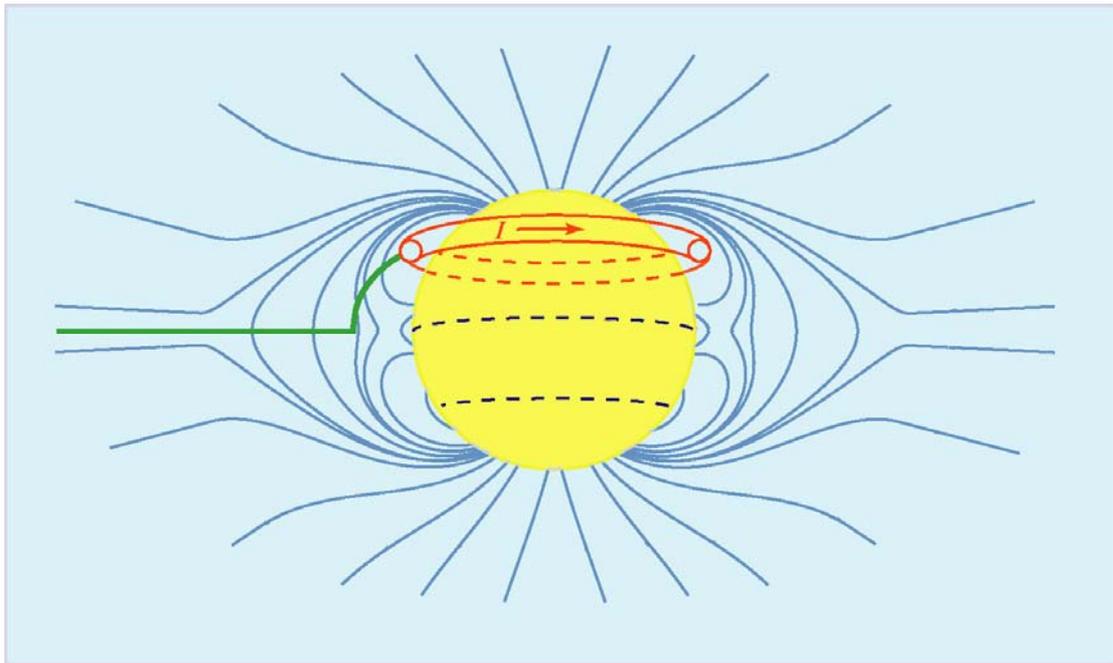

Figure 6. Field lines of the solar global magnetic field typical for minimum activity adopted in the axially-symmetric flux rope model (Filippov et al., 2002). The dashed lines show the polarity inversion lines at the photospheric surface, and the green line, the possible equilibrium positions of the flux rope.

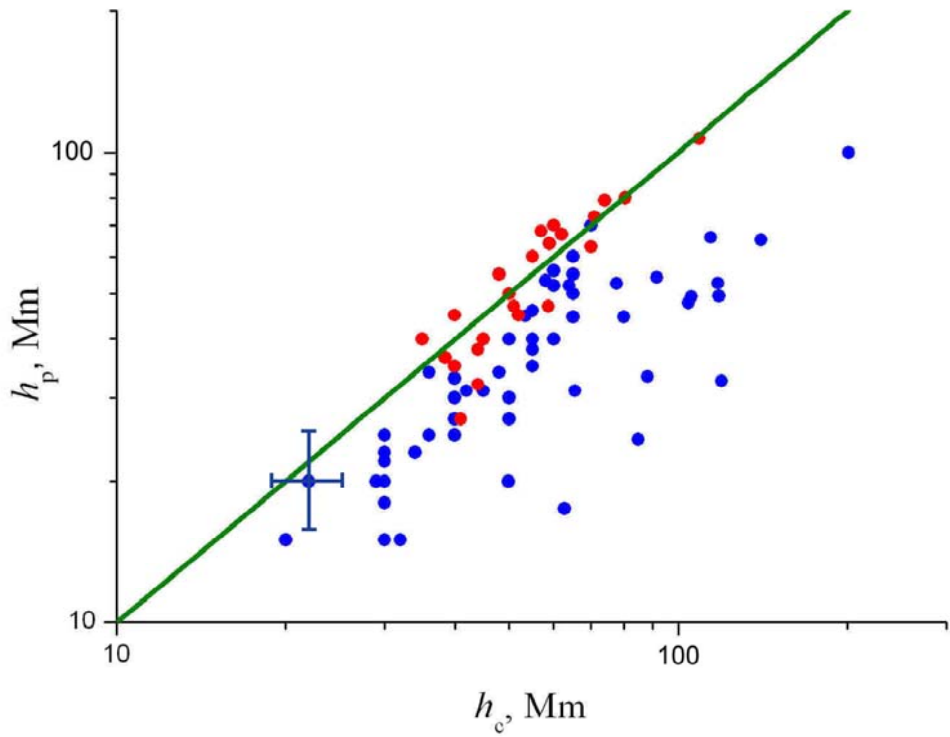

Figure 7. The observed filament height above the chromosphere $h_p$ versus the critical height of stable filament equilibrium $h_c$. The blue circles correspond to the filaments which safely passed the west limb. The red circles correspond to the filaments which disappeared from the disk. The straight green line corresponding to an equality of these quantities is the stability boundary.

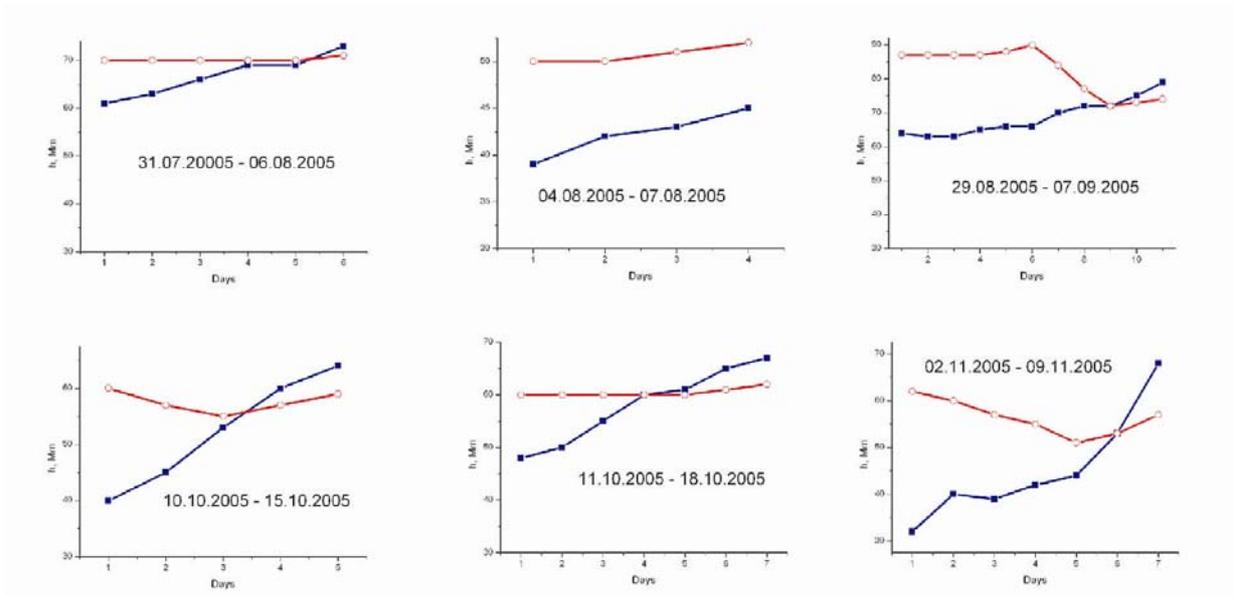

Figure 8. Temporal behaviour of the filament height $h_p$ (blue squares) above the photosphere and the critical height $h_c$ (open red circles) several days before eruption in the latter part of 2005.